\documentclass{aa}
\usepackage{graphicx}
\usepackage{txfonts}

\begin{document}
   \title{Visible spectroscopy in the neighborhood of 2003EL$_{61}$}
   \author{N. Pinilla-Alonso
          \inst{1}
	  J. Licandro
	  \inst{2}
	  \and	  
	  V. Lorenzi
	  \inst{1}}

   \offprints{N. Pinilla-Alonso}

  \institute{Fundaci\'on Galileo Galilei \& Telescopio Nazionale Galileo, P.O.Box 565, E-38700, S/C de La Palma, Tenerife, Spain.
              \email{npinilla@tng.iac.es}
              \and
	     Instituto de Astrof\'{\i}sica de Canarias, c/V\'{\i}a L\'actea s/n, E38205, La Laguna, Tenerife, Spain.\\}
   \date{Received ; accepted July2008}

% \abstract{}{}{}{}{} 
% 5 {} token are mandatory
   \abstract
   {The recent discovery of a group of trans-neptunian objects (TNOs) in a narrow region of the orbital parameter space and with surfaces composed of almost pure water ice, being 2003 EL$_{61}$  its largest member, promises new and interesting results about the formation and evolution of the TNb and the outer Solar System.}
   {The aim of this paper is to obtain information of the surface properties of two members of this group ((24835) 1995 SM$_{55}$, (120178) 2003 OP$_{32}$) and three potential members (2003 UZ$_{117}$, (120347) 2004 SB$_{60}$ and 2005 UQ$_{513}$) and to use that in order to confirm or reject their association.}
   {We obtained visible spectra of five TNOs using the 3.58m Telescopio Nazionale Galileo at the ``Roque de los Muchachos Observatory'' (La Palma, Spain)}
   {The spectra of the five TNOs are featureless within the uncertainties and with colors from slightly blue to red ($-2 < S' < 18$\%$/0.1\mu$m). No signatures of any absorption are found.}
   {We confirm the association of 1995 SM$_{55}$ and 2003 OP$_{32}$ with the group of 2003 EL$_{61}$ as their spectra are almost identical to that of 2003 EL$_{61}$. Only one of the three candidates, 2003 UZ$_{117}$,  can be considered as a possible member of  the EL$_{61}$-group, as its visible spectrum is compatible with a spectrum of a surface composed of almost pure water ice and no complex organics. The other two, 2004 SB$_{60}$ and 2005 UQ$_{513}$ are red and must be considered as interlopers.}
   
   \titlerunning{Visible spectroscopy next to 2003 EL$_{61}$}
   \maketitle

%________________________________________________________________

\section{Introduction}

TNO 2003 EL$_{61}$  is the largest member of a group of TNOs (hereafter EL$_{61}$-group) with orbits in a narrow region of the orbital parameter space (41.6 $<$a$<$ 43.6 AU, 25.8 $<$i$<$ 28.2 deg., 0.10 $<$e$<$ 0.19) (Pinilla-Alonso et al., \cite{Pinilla-AlonsoRR43}) and with surfaces composed of almost pure water ice what means: (a) The visible is featureless within the S/N. There is no clear evidence of any absorption reported for other TNOs; (b) The visible is neutral ; (c) It presents two deep absorption bands centered at 1.5 and 2.0 $\mu$m, indicative of water ice and compatible with the presence of the crystalline phase. Brown et al. (\cite{Brownfamily}) claim that this group is a family of fragments product of a giant collision happened in the trans-neptunian belt (TNb).

The first identified members of the group are: (136108) 2003 EL$_{61}$, (55636) 2002 TX$_{300}$, (145453) 2005 RR$_{43}$, (120178) 2003 OP$_{32}$, (19308) 1996 TO$_{66}$, and (24835) 1995 SM$_{55}$. The identification of the members of this group is possible thanks to the study of their spectra as all of them show the same characteristics. Near infrared spectra have been published for all these objects and they show very deep water ice absorption bands (Pinilla-Alonso et al. \cite{Pinilla-AlonsoRR43} and references therein). Visible spectra have been published only for 2003 EL$_{61}$, 2002 TX$_{300}$, and 2005 RR$_{43}$ (Tegler et al. \cite{TegEL61}, Licandro et al. \cite{LicTX}, Pinilla-Alonso et al., \cite{Pinilla-AlonsoRR43}) and are all similar, featureless with an almost neutral slope.

Spectroscopy is the best way to study the surface of these objects but for those that are too faint, photometry is a usefull tool. Visible colors of objects 2003 OP$_{32}$, 1995 SM$_{55}$ and 1996 TO$_{66}$ are indicative of a neutral spectral slope in this spectral region (MBOSS database http://www.sc.eso.org/~ohainaut/MBOSS/). Another object, (86047) 1999 OY$_{3}$ has been identified as a member of the group as its visible and near-infrared colors are compatible with the characteristics of a typical spectrum of a member of the group (%\textbf{refNoll}
Ragozzine et al. \cite{Ragozzine} and Hainaut \& Delsanti \cite{MBOSS}).\\

Fitting scattering models to the visible and near-infrared spectra of 2005 RR$_{43}$ and 2003 EL$_{61}$ Pinilla-Alonso et al. (\cite{Pinilla-AlonsoRR43} and \cite{Pinilla-AlonsoEL61}) show that their surfaces are composed of a mixture of amorphous and crystalline water ice and discard the presence of a significant amount of other components. In particular they discard the presence of complex organics typically red in color produced by the irradiation of hydrocarbons and/or alcohols. Furthermore, they conclude that 2003 EL$_{61}$ (and probably the other members of the group) has a significant smaller fraction of carbon chains on its surface than the other TNOs. This carbon-depleted population of TNOs is located in an unstable region of the TNb crossed by resonances and could be the source of some of the carbon-depleted comets already noticed by A'Hearn et al. (\cite{AHearn95}). This possibility makes this group particularly intriguing.

Pinilla-Alonso et al. (\cite{Pinilla-AlonsoEL61}) also show that, if the surface of 2003 EL$_{61}$ is the product of a large collision, the relative percentage of amorphous and crystalline ice obtained from the models, implies that this collision should have happened more than 10$^{8}$ years ago, in agreement with dynamical predictions (Ragozzine \& Brown \cite{Ragozzine}) and laboratory experiments (Zheng et al. \cite{ZenghWater}).

Assuming that this population is the product of a giant collision, Ragozzine \& Brown (\cite{Ragozzine}) applied techniques developed for the study of the asteroid families to the study of the objects in the vicinity of 2003 EL$_{61}$. They compute the minimum ejection velocity of potential fragments required to change their orbital parameters to their actual values ($\Delta$v$_{min}$). In a second approximation they also consider diffusion in resonances ($\delta$v$_{min}$, see  Ragozzine \& Brown \cite{Ragozzine} for details). They find that all the objects in the group can be explained by a velocity dispersion of 150 m s$^{-1}$ from a single collision location and considering diffusion in eccentricity in resonances. From these dynamical considerations, they even identify a group of TNOs that are potential candidates to be part of the group.

The study of the surface properties of these objects can be used to identify other possible members of the EL$_{61}$-group and to identify other TNOs that, having similar dynamical properties, are not members, this is important to our understanding of the origin of EL$_{61}$-group. 

In this paper, we present visible spectra of three of the candidates in Ragozzine \& Brown (\cite{Ragozzine}), 2003 UZ$_{117}$, (120347) 2004 SB$_{60}$ and 2005 UQ$_{513}$, and of two of the already known members of the EL$_{61}$ group without published visible spectroscopy, (24835) 1995 SM$_{55}$, (120178) 2003 OP$_{32}$.

\section{Observations}

\begin{table*}
\centering
\begin{tabular}{l c c c c c c c}
\hline
\hline

Object & Date & r & Delta & phase & n & T$_{exp}$ & airmass\\ \hline\hline

1995 SM$_{55}$ & 18.11-18.20 & 38.805 & 38.114 & 1.1 & 3 & 2400 & 1.01-1.06\\  \hline
1995 SM$_{55}$ & 18.21-18.24 & 38.805 & 38.114 & 1.1 & 1 & 1800 & 1.01\\ \hline
2003 OP$_{32}$ & 15.89-15.96 & 41.265 & 40.411 &  0.7 & 3 & 1800 & 1.11-1.21\\ \hline  
2003 UZ$_{117}$ & 17.07-17.22 & 39.461 & 38.859 & 1.2 & 4 & 2400 & 1.51-1.11\\ \hline
2003 UZ$_{117}$ & 17.22-17.24 & 43.892 & 42.968 & 0.5 & 1 & 1800 & 1.11 \\ \hline
2004 SB$_{60}$ & 16.87-16.99 & 43.892 & 42.968 & 0.5 & 5 & 1800 & 1.04-1.45\\  \hline
2005 UQ$_{513}$ & 16.00-16.06 & 48.859 & 47.996 & 0.6 & 2 & 2400 & 1.00-1.07 \\  \hline
2005 UQ$_{513}$ & 16.15-16.22 & 48.859 & 47.996 & 0.6 & 2 & 2400 & 1.22-1.50\\  \hline
2005 UQ$_{513}$ & 16.22-16.24 & 48.859 & 47.996 & 0.6 & 1 & 1600 & 1.56-1.70\\  \hline
2005 UQ$_{513}$ & 17.03-17.05 & 48.859 & 47.990 & 0.6 & 1 & 1800 & 1.04-1.01\\ \hline\hline

\end{tabular}
  \caption{Observations from 15 to 18 Sept 2007. Date: day and time of the beginning and the end of the expossure; r: distance from the sun to the object (AU); Delta: distance from the observer to the object (AU); phase: phase angle of the object ($^{\circ}$); n: number of individual exposures; T$_{exp}$: expossure time of each single spectrum measured in seconds; Airmass: airmass range.}
  \label{Table1}
\end{table*}

Visible spectra of TNOs 1995 SM$_{55}$, 2003 OP$_{32}$, 2003 UZ$_{117}$, 2004 SB$_{60}$ and 2005 UQ$_{513}$, were done with the 3.58m Telescopio Nazionale Galileo (TNG, El Roque de los Muchachos Observatory, Canary Islands, Spain) on three consecutive nights from September 15 to 17, 2007. The spectrograph DOLORES with the LR-R grism and the 2.0"  slit width was used. Several spectra of each object with expossure times from 1600 to 2400 sec, covering the  0.52 to 0.86$\mu $m spectral range, were obtained (see Table \ref{Table1}). Each object was shifted in the slit by 5'' between consecutive spectra to better correct the fringing. 

Images were bias corrected using the over-scan region and flat-field corrected using lamp flats. The two-dimensional spectra were extracted, sky background subtracted, and collapsed to one dimension. The wavelength calibration was done using the Neon and Argon lamps.  All the spectra of each TNO, obtained at different positions in the slit, were averaged and rebined to obtain a higher S/N. 

\begin{figure}
	\centering
	\includegraphics[width=\columnwidth]{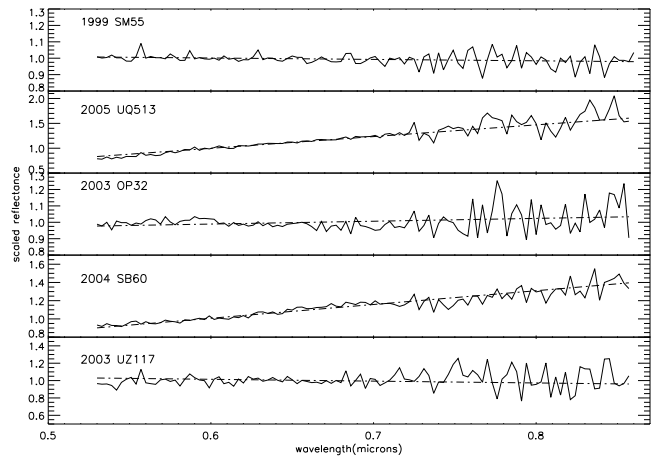}
	\caption{Visible Spectra normalized to unity at 0.55 $\mu $m}
 	\label{Fig1}
 \end{figure}

To correct for telluric absorption and to obtain the relative reflectance, the G stars Landolt (SA) 110-36, Landolt (SA) 115-271, Landolt (SA) 93-101, Landolt (SA) 112-13, Landolt (SA) 98-978, Landolt (SA) 110-113 and Landolt (SA)115-27  (Landolt, \cite{Landolt}) were observed at different airmasses (similar to those of the TNOs) during each night, before and after the TNO observations, and used as solar analogue stars. The spectrum of each object was divided by those of the solar analogue stars observed the same night and at similar airmasses, and then normalized to unity around 0.55$\mu $m thus obtaining the normalized reflectance. The obtained spectra are shown in Fig. \ref{Fig1}.

\section{Analysis of the spectra}

The spectra of the five TNOs are featureless within the uncertainties. No signatures of any absorption band centered at 0.7 $\mu$m, typically observed in low albedo main belt asteroids and attributed to silicate aqueous alteration, are found. This very weak absorption has been reported for other TNOs (e.g Lazzarin et al. \cite{LazzPhyll}; Fornasier et al. \cite{FornaPhyll}).

However, the spectra show variety in color, from bluish to red. To perform a quantitative analysis of the color distribution of the EL$_{61}$-group and to discriminate if the candidates studied belong to it, we computed the spectral gradient $S'[\% (0.1\mu$m)$^{-1}]$ as defined by Jewitt. (\cite{Jewittslope}):

\begin{equation}
S'=\frac{\delta S}{\delta\lambda}\times {\overline{S}^{-1}}
\end{equation}

where $S$ is the relative reflectance of the object and $\overline{S}$ is the mean value of this reflectance in the wavelength range over which ${dS}$/${d\lambda}$ is computed. $S'$ is a  measurement of the variation of $S(\lambda)$ over an interval $\Delta\lambda$. Assuming that a featureless spectrum has an overall linear shape (as a first order approach), we made a linear fitting of the visible spectra normalized at 0.55 $\mu$m, obtaining $S'$. Results are shown in Table \ref{Table2}, together with the dynamical parameters of each TNO. 

The analysis of the solar analogue stars observed the three nights demonstrate that uncertainties up to 2\%$/0.1\mu$m in $S'$ are possible due to systematic errors. This is smaller than the uncertainties due to spectrophotometric determinations that can easily be larger than 5\%$/0.1\mu$m.

\begin{table*}
\centering
\begin{tabular}{l c c c c c c c}
\hline\hline

Object & a & e & i & q & $\Delta\upsilon_{min}$  & $\delta$v$_{min}$ & \textit{S'}\\ \hline\hline
1995 SM$_{55}$ & 41.685 & 0.102 &27.1 &37.435 & 149.7 & 123.3 & -0.9 $\pm$ 2.0 \\  \hline
2003 OP$_{32}$ & 43.366 & 0.109 &27.2 &38.658 & 123.3 & 91.4 & 1.7 $\pm$ 2.0 \\  \hline
2003 UZ$_{117}$& 44.062 &0.128  &27.5 &38.411 & 66.8 & 60.8 & -2.1 $\pm$ 2.0 \\  \hline
2004 SB$_{60}$ & 42.082 &0.108  &23.9 &37.536 & 221.0 & 218.5 & 12.6 $\pm$ 2.0\\  \hline
2005 UQ$_{513}$& 43.291 &0.151 &25.7 &36.763 & 199.2 & 39.0 & 18.1 $\pm$ 2.0 \\  \hline\hline

\end{tabular}
\caption{The orbital elements $a$, $e$, $i$, and $q$ (semi-major axis [AU], eccentricity, inclination [$^{\circ}$] and perihelion distance [AU]); $\Delta$v$_{min}$ and $\delta$v$_{min}$ [m s$^{-1}$] (velocities from Ragozzine \& Brown (\cite{Ragozzine}); $S'$ the spectral gradient of five TNOs presented in this paper.}
 \label{Table2}
\end{table*}

\subsection{Family members: 1995 SM$_{55}$, 2003 OP$_{32}$}

TNOs 1995 SM$_{55}$, 2003 OP$_{32}$ are already considered as members of the family. Spectral gradients computed from their spectra (see table 2.) are consistent with Ragozzine \& Brown (\cite{Ragozzine}) values obtained from photometry 1.79 $\pm$ 2.60 and -1.09 $\pm$ 2.20 \%$/0.1\mu$m respectively. 

The spectral gradients computed from the spectra of the other members of the group 2003 EL$_{61}$, 2002 TX$_{300}$, and 2005 RR$_{43}$ (Pinilla-Alonso et al. \cite{Pinilla-AlonsoEL61}, Licandro et al. \cite{LicTX}, Pinilla-Alonso et al., \cite{Pinilla-AlonsoRR43}) are  0.0, 1.0 and 0.4 respectively. These give a mean $S'$=0.4 $\pm$ 1.0\%$/0.1\mu$m for the EL$_{61}$-group.

\subsection{Candidates: 2003 UZ$_{117}$, 2004 SB$_{60}$ and 2005 UQ$_{513}$}

The visible spectra of the three observed TNOs candidates to belong to the EL$_{61}$-group according to Ragozzine \& Brown (\cite{Ragozzine}), are all featureless but show variety in colors. Two of them (2004 SB$_{60}$ and 2005 UQ$_{513}$) are too red ($S'=$ 12.6 and 18.1 $\pm$ 2.0 \%$/0.1\mu$m respectively) to be members of the group. The surfaces of these objects probably contain a significant fraction of processed materials composed of complex organics.

On the other hand, 2003 UZ$_{117}$ should be considered as a strong candidate to be part of the EL$_{61}$-group. Its spectrum is featureless in the visible and bluish with an spectral gradient computed from the spectrum, $S'$=-2 $\pm$ 2 \%$/0.1\mu$m, close to the mean value for the whole group and compatible with the spectral gradient computed from the photometry (Ragozzine et al. \cite{Ragozzine}).

This fact is consistent, but not conclusive as we need near-infrared photometry and/or spectroscopy to confirm the presence of water ice. It is well known that solar irradiation during Gy. sublimates volatiles from the surface of trans-neptunians (Gil-Hutton. \cite{Gil-Hutton}) and covers it with neutral and low albedo mantles, so some TNOs exist with solar colors in the visible but weak or null water ice presence on their surface (ex. Orcus, de Bergh et al. \cite{DeberghOrcus}). Consequently, we confirm 2003 UZ$_{117}$ is a strong candidate to belong to the family but we emphasize we need further studies that should confirm a high abundance of water ice and the absence of a significant amount of complex organics on its surface.

From the dynamical point of view, considering the orbital parameters and spectral properties, 2005 UQ$_{513}$ is a good candidate to be a collisional fragment of 2003 EL$_{61}$, much better than 2004 SB$_{60}$. Considering diffusion, 2005 UQ$_{513}$ has one of the lowest minimum ejection velocities ($\delta$v$_{min}$ = 39.0 km/h), while $\delta$v$_{min}$ = 218.5 for 2004 SB$_{60}$. The spectra of both TNOs are incompatible with those of the rest of the EL$_{61}$-group and it is very probable that traces of organics and/or rests of volatiles responsible for the red color of the visible spectra can be present in the near-infrared spectra of these objects.

\begin{figure}
	\centering
	\includegraphics[width=\columnwidth]{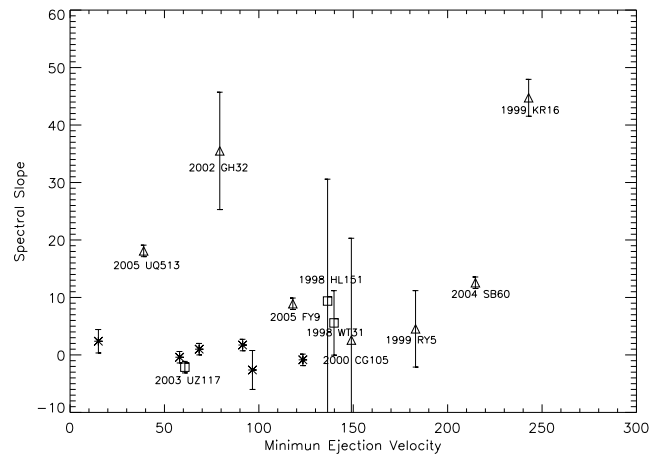}
	\caption{Spectral Gradient versus minimum ejection velocity considering diffusion in resonances. We represent with asterism objects in the family. Four objects with $\delta$v$_{min} \leq $ 150 are interlopers (represented by triangles). Three other objects with $\delta$v$_{min} \leq $ 150 need more observations to be considered as candidates or discarded (represented by squares).}
 	\label{Fig2}
 \end{figure}

In Table \ref{Table3}, we present the dynamical parameters and $S'$ of all the other TNOs (members and candidates) related to the EL$_{61}$-group with spectroscopic and/or spectrophotometric data already published. In Figure \ref{Fig2} we present a plot of the $S'$ vs $\delta$v$_{min}$ of all TNOs in Tables \ref{Table2} and \ref{Table3}. Represented by asterisks are the confirmed members of the EL$_{61}$-group, by open triangles those TNOs that due to their spectral properties, either in the visible or in the near-infrared, are non-members, and by open square the TNOs with an $S'$ that could be $<$ 3\%$/0.1\mu$m considering the uncertainties.
Notice that there are 13 TNOs with $\delta$v$_{min} < 150$m/s, as we mentioned, the minimum velocity required to explained the actual distribution in orbital parameters of the members of the group, allowing diffusion in eccentricity in resonances. 6 of them are confirmed members (46\%), 4 non-members (31\%) and 3 candidates (23\%). There are still 4 candidates with $\delta$v$_{min} < 150$m/s that have not yet been observed either photometrically or spectrophotometrically.

\begin{table*}
\centering
\begin{tabular}{l c c c c c c r c}
\hline\hline

Object & a & e & i & q & $\Delta\upsilon_{min}$  & $\delta$v$_{min}$ & \textit{S'} & Ref \\ \hline\hline
1996 TO$_{66}$ & 43.283 & 0.121 &27.4 &38.051 & 24.2 & 15.0 & 2.38 $\pm$ 2.04 & (3)\\  \hline
2005 RR$_{43}$ & 43.115 &0.138  &28.6 &37.175 & 111.2 & 58.0 & -0.4 $\pm$ 2.0 & (4)\\  \hline
2003 TX$_{300}$& 43.216 &0.125  &25.8 &37.821 & 107.5 & 68.4 & 1 $\pm$ 2.00 & (1)\\  \hline
2002 GH$_{32}$& 42.185 & 0.086 & 26.6 & 38.541 & 141.9 & 79.3& 35.25 $\pm$ 10.21 & (3)\\ \hline
2003 EL$_{61}$ & 43.240 & 0.192  & 28.2 & 34.931 & 323.5 && 0.0 $\pm$ 2.0 & (5)\\  \hline
1999 OY$_{3}$ & 42.185 & 0.086 & 26.6&3 8.541 & 292.8 & 96.6 & -2.62 $\pm$ 3.39 & (3)\\  \hline
2005 FY$_{9}$ & 45.543 & 0.158 & 29.0 & 38.333 & 141.2 & 118.0 & 8.9 $\pm$ 1.0 & (2)\\  \hline
1998 HL$_{151}$& 40.930 & 0.091 & 28.0 & 37.224 & 142.5 & 136.4 & 9.83 $\pm$ 21.1& (3) \\ \hline
1998 WT$_{31}$& 45.821 & 0.180 & 28.7 & 37.581 & 233.3 & 139.8 & 5.57 $\pm$ 5.61 & (3)\\ \hline
2000 CG$_{105}$& 46.280 & 0.038 & 28.0 & 44.533 &  & 149.0 & 2.58 $\pm$ 17.72 & (5)\\ \hline
1999 RY$_{215}$& 45.228 & 0.236 & 22.2 & 34.554 & - & 183.0 & 4.54 $\pm$ 6.65 & (3)\\ \hline
1999 KR$_{16}$& 49.206 & 0.309 & 24.8 & 34.001 & - & 242.9 & 44.74 $\pm$ 3.21 & (3)\\ \hline

\end{tabular}
\caption{The orbital elements $a$, $e$, $i$, and $q$ (semi-major axis[AU], eccentricity, inclination [$^{\circ}$] and perihelion distance [AU]), the $\Delta$v$_{min}$ and $\delta$v$_{min}$ [m s$^{-1}$] quantities computed by Ragozzine \& Brown (\cite{Ragozzine}) and the spectral slope $S'$ of other members and candidates of the EL$_{61}$-group. References in column (8) are for the measured spectral gradient: (1) Licandro et al. (\cite{LicTX}), (2) Licandro et al. (\cite{LicFY9}), (3) Ragozzine \& Brown (\cite{Ragozzine}) and references therein, (4) Pinilla-Alonso et al. (\cite{Pinilla-AlonsoRR43}) and (5) Pinilla-Alonso et al. (\cite{Pinilla-AlonsoEL61})}
  \label{Table3}
\end{table*}

\section{Conclusions.}

We present visible spectra of 5 TNOs in the neighborhood of 2003 EL$_{61}$ obtained with the 3.58m Telescopio Nazionale Galileo (La Palma, Spain). Two of them are members of the EL$_{61}$-group (1995 SM$_{55}$, 2003 OP$_{32}$) and do not have previous published visible spectroscopy. The other three are potential group members (2003 UZ$_{117}$,2004 SB$_{60}$ and 2005 UQ$_{513}$) according to Ragozzine \& Brown (\cite{Ragozzine}).

The spectra of the five TNOs are featureless within the uncertainties, and with colors from slightly blue to red ($-2 < S' < 18$\%$/0.1\mu$m). No signatures of any absorption band are found.

Only one of the three candidates, 2003 UZ$_{117}$, can be considered as a possible member of the EL$_{61}$-group considering that its visible spectrum is compatible with a spectrum of a surface composed of almost pure water ice and no significant amount of complex organics, but near-infrared photometry or spectroscopy is needed to confirm the presence of water ice.
The other two, 2004 SB$_{60}$. and 2005 UQ$_{513}$  are red ($S'$= 13 and 18\%$/0.1\mu$m respectively) thus they cannot be members of the EL$_{61}$-group. The surface of these objects probably contains a significative fraction of complex organics.

\begin{acknowledgements}
Based on observations made with the Italian Telescopio Nazionale Galileo (TNG) operated on the island of La Palma by the Fundaci\'on Galileo Galilei of the INAF (Istituto Nazionale di Astrofisica) at the Spanish Observatorio del Roque de los Muchachos of the Instituto de Astrof\'isica de Canarias.
\end{acknowledgements}

\end{document}